\documentstyle[eqsecnum,prd,aps,epsf]{revtex}
\begin{document}
 \makeatletter
 \newdimen\ex@
 \ex@.2326ex
 \def\dddot#1{{\mathop{#1}\limits^{\vbox to-1.4\ex@{\kern-\tw@\ex@
  \hbox{\tenrm...}\vss}}}}
 \makeatother
\thispagestyle{empty}
{\baselineskip0pt
\leftline{\large\baselineskip16pt\sl\vbox to0pt{\hbox{\it Department of Physics}
               \hbox{\it Osaka City  University}\vss}}
\rightline{\large\baselineskip16pt\rm\vbox to20pt{\hbox{OCU-PHYS-176}
            \hbox{OU-TAP 139}
            \hbox{WU-AP/105/00}
            \hbox{\today} 
\vss}}%
}
\vskip3cm
\begin{center}{\large 
{\bf Newtonian Analysis of Gravitational Waves from Naked Singularity
}}
\end{center}
\begin{center}
 {\large Ken-ichi Nakao$^{1}$, Hideo Iguchi$^{2}$ and Tomohiro
 Harada$^{3}$
} \\
{\em $^{1}$Department of Physics,~Osaka City University} \\
{\em Osaka 558-8585,~Japan}\\
{\em $^{2}$Department of Earth and Space Science, Graduate School of 
Science, Osaka University} \\
{\em Toyonaka, Osaka 560-0043, ~Japan}\\
{\em $^{3}$Department of Physics,~Waseda University} \\
{\em Oh-kubo, Shinjuku-ku, Tokyo 169-8555,~Japan} \\
\end{center}
\begin{abstract}
 
Spherical dust collapse generally forms a 
shell focusing naked singularity at the symmetric center. 
This naked singularity is massless. Further the Newtonian gravitational 
potential and speed of the dust fluid elements are 
everywhere much smaller than unity until the central shell focusing 
naked singularity formation 
if an appropriate initial condition is set up. 
Although such a situation is highly relativistic, the analysis by the 
Newtonian approximation scheme is available even in the vicinity 
of the space-time singularity. This remarkable feature 
makes the analysis of such 
singularity formation very easy. We investigate non-spherical even-parity 
matter perturbations in this scheme by complementary using 
numerical and semi-analytical approaches, and estimate linear gravitational 
waves generated in the neighborhood of the naked singularity by 
the quadrupole formula. The result shows good agreement with 
the relativistic perturbation analysis recently performed by 
Iguchi et al. The energy flux of the gravitational waves  
is finite but the space-time curvature carried by them  
diverges. 

\vskip0.5cm
\noindent
PACS number(s): 04.25.Nx,04.30.Db,04.20.Dw

\end{abstract}

\section{Introduction}
\label{sec:intro}
General relativity predicts that gravitational collapse of 
massive objects leads to space-time singularities in rather general 
circumstances in our universe\cite{penrose65,hawking67,hp70} 
and such singularities might be accompanied 
with the blow up of physical quantities (energy density, pressure and 
curvature of space-time, etc). 
The known physical laws including general relativity 
itself will break down in the neighborhood of 
the space-time singularity and hence 
the quantum theory of gravity is believed to be necessary to 
describe the physical phenomena in such a region. 

One of the important issues is whether the space-time singularities 
formed in our universe are visible or not for the 
observer (us) far from the region where the gravitational 
collapse occurs. 
The cosmic censorship conjecture proposed by Penrose gave a strong motivation 
to investigate this problem\cite{penrose69}. 
Roughly speaking, this conjecture states that the singularity 
is not visible for any observer if it is resulted from physically 
reasonable initial conditions. However, this conjecture 
has not yet been proven. 
Rather candidates of the counterexample of this conjecture 
have been found. 

The simplest example is the gravitational collapse of spherically 
symmetric dust fluid. 
There is an exact solution of Einstein equations for this 
case; the so-called Lema\^{\i}tre-Tolman-Bondi (LTB) solution. 
After Eardley and Smarr have pointed out the occurrence  
of a central shell focusing naked singularity 
of marginally bound dust collapse\cite{es1979}, several 
theoretical efforts 
revealed the genericity of the shell focusing naked singularity 
formation in the LTB
solution\cite{christodoulou1984,newman1986,jd1993,sj1996,jjs1996}. 
There are several researches for more  
general spherically symmetric system and those revealed 
that non-vanishing pressure does not necessarily 
prevent the formation of the central shell focusing naked 
singularity\cite{op1988,op1990,harada1998,hin1998}. 

On the other hand, there are not so much researches for non-spherical 
systems as the spherically symmetric cases since 
general analytic approach is impossible. 
Nakamura et al. and Nakamura and Sato performed numerical simulations 
for axisymmetric perfect fluid system\cite{nmms82,ns82}. 
Their results suggest that 
naked singularities might be formed if the initial internal energy 
of the fluid elements is very small and initial configuration 
is sufficiently elongated. This result is consistent with 
the hoop conjecture proposed by Thorne, which states that 
black holes with horizons form when and only when a mass $M$ 
gets compacted into a region whose circumference in every 
direction is $C\gtrsim4\pi M$\cite{Ref:HOOP}.
Shapiro and Teukolsky also performed numerical simulations
for the axisymmetric collision-less-particle system and showed a 
possibility of naked singularity formation which is also consistent with the 
hoop conjecture\cite{st91}. The critical behavior of the axisymmetric 
gravitational waves is also a candidate of the naked singularity 
formation\cite{ab1993}. An example which can be investigated analytically 
is the shell focusing naked singularity formation in the Szekeres 
solution\cite{jk96}. Anyway, even in the non-spherical cases, there are 
several examples of naked singularity formation. 

There might be readers who state that numerical examples 
can not be regarded as candidates of naked singularity formation 
since the numerical simulation can not reveal the 
global structure of the space-time and further can not deal with 
infinite quantities (the numerical technique recently proposed by 
H${\ddot {\rm u}}$bner 
might be able to avoid this difficulty\cite{hubner}). 
However, the present authors would like to 
stress that if the region of sufficiently high energy, large pressure and 
large space-time curvature is visible from an observer far from 
the region, it should be regarded as a naked singularity. 
The region with extremely large physical quantities 
will be described by the quantum theory of gravity which is little 
known and hence such region is equivalent 
to a space-time singularity in practical sense. 

An important point of view for the naked singularity formation process 
has been proposed by Nakamura et al.\cite{nsn1993}. 
They pointed out that the visible strong curvature region 
may be a candidate of strong gravitational wave sources. 
Gravitational waves generated in such a region will propagate out 
from there since there is no event horizon. 
From this point of view, Chiba investigated cylindrical 
gravitational collapse but his result is not consistent to the 
Nakamura et al's conjecture\cite{chiba96}. The numerical simulations by 
Shapiro and Teukolsky essentially agrees with Chiba's result 
with respect to the gravitational radiation. 
However, we should be concerned about the numerically covered domain 
since the gravitational waves might be emitted 
form just the neighborhood of the naked singularity. 
Recently, present authors investigated the behavior of aspherical 
perturbations in the LTB space-time and showed that the Weyl 
curvature of even mode perturbations corresponding to 
outgoing gravitational waves diverges\cite{ihn1999b}. 
Although this result was obtained by numerical integration 
of linearized Einstein equations, the numerical stability 
is guaranteed much better than that of the numerical simulation of 
full Einstein equations. Further, linearized Einstein equations 
were solved as the characteristic initial value problem and hence 
the numerically covered domain 
of the space-time by this analysis is much wider than the 
previous two numerical 
simulations. The results of the linear perturbation analysis 
of the LTB space-time suggest that gravitational 
waves will be emitted in the formation process 
of naked singularities. 

The purpose of the present paper is to re-analyze the 
dynamics of perturbations of the LTB space-time in the framework of the 
Newtonian approximation. In order that the singularity of the 
spherically symmetric space-time is naked, 
``the gravitational potential'' $2M/R$ is smaller than unity in 
the neighborhood of the singularity, where $M$ is the Misner-Sharp mass 
function and $R$ is the areal radius. The central shell focusing 
naked singularity 
of the LTB space-time satisfies this condition and further 
the gravitational potential vanishes even at this singularity. 
The speed of the dust fluid is also much smaller than the speed of light 
before and at the central shell focusing naked singularity formation. 
Therefore the Newtonian approximation seems to be available, 
even though the space-time curvature is infinite at the singularity. 
The advantage of the Newtonian approximation scheme is that 
the dynamics of perturbations of the dust fluid and 
gravitational waves generated by the motion of the dust fluid 
are separately estimated; the evolution of the perturbations 
of the dust fluid are obtained by the Newtonian dynamics and 
the gravitational radiation is obtained by the quadrupole formula. 
Hence the semi-analytic estimate of the gravitational radiation 
due to the matter perturbation of the LTB space-time is possible 
if we adopt the Newtonian approximation.  
The results of the Newtonian perturbation analysis well agrees 
with the relativistic perturbation analysis by the present authors. 
This suggests that the Newtonian analysis will be a powerful tool 
in the analysis of some category of naked singularity. 
However, we should stress that the neighborhood 
of the naked singularity is not Newtonian in ordinary sense, 
because there is indefinitely strong tidal force.   
Further, studies of quantum effects have revealed that the violent 
quantum particle creation occurs in this 
space-time\cite{bsvw1998a,bsvw1998b,vw1998,hin2000a,hin2000b}. 
The particle creation is just a highly relativistic phenomenon. 
The Newtonian approximation scheme is available to describe the dynamics 
of the neighborhood of the naked singularity but the situation 
is not Newtonian. 

This paper is organized as follows. In Sec.2, we briefly 
review the LTB solution. In Sec.3, we consider 
the Newtonian approximation of a spherically symmetric dust 
collapse and show the relation between the Eulerian, Lagrangian 
and synchronous-comoving (SC) coordinate systems in this approximation 
scheme. Further in this section, we show the validity of the Newtonian 
approximation scheme even in the neighborhood of the central shell 
focusing naked singularity as long as we adopt the Eulerian coordinate 
system. The basic equations for the even mode of perturbations 
are presented in Sec.4. In order to estimate the gravitational 
radiation generated by aspherical perturbations of dust fluid, 
we need the knowledge about mass-quadrupole moment. 
We show an explicit expression of the
quadrupole moment in Sec.5. Then we show the numerical calculation 
in Sec.6 and the asymptotic behavior of the mass-quadrupole 
moment is presented in Sec.7. 
Finally, Sec.8 is devoted for summary and discussion. 

In this article, we adopt $G=c=1$ unit and basically follow the convention 
and notation in Ref.\cite{wald}. The Greek indices mean components 
of tensors, while the Latin indices except for $\ell$ and $m$ 
represent a type of a tensor. We use the sub- or super-scripts 
$\ell$ and $m$ to denote spatial components of tensors. 

\section{The Spherically Symmetric Dust Collapse}

In general, the gravitational collapse of a spherical dust ball produces a
shell focusing naked singularity at the symmetric center.
This singularity can be locally or globally naked in accordance with
the initial rest-mass density configuration.
The solution of the Einstein equations describing such a situation
is known as the LTB space-time. The line element
is given by
\begin{equation}
ds^{2}=-dt^{2}+{(\partial_{r}R)_{t}^{2}\over
1+f(r)}dr^{2}+R^{2}d\Omega^{2}, \label{eq:SC-line}
\end{equation}
where $d\Omega^{2}$ is the line element of an unit 2-sphere.
The stress-energy tensor of the dust fluid is given by
\begin{equation}
T_{ab}={\bar \rho}{\bar u}_{a}{\bar u}_{b},
\end{equation}
where ${\bar \rho}$ is the rest-mass density and 
${\bar u}_{a}$ is the 4-velocity of the dust fluid element. 
In the coordinate system (\ref{eq:SC-line}), the components of the 
4-velocity ${\bar u}^{\mu}$ is given by
\begin{equation}
{\bar u}^{\mu}=(1,0,0,0).
\end{equation}
Einstein equations and the equation of motion for the dust fluid
lead to the equations for the areal radius $R$ and rest-mass density 
${\bar \rho}$ as 
\begin{eqnarray}
(\partial_{t}R)_{r}^{2}&=&f(r)+{F(r)\over R}, \label{eq:energy-eq}\\
F(r)&=&8\pi\int_{0}^{r}{\bar \rho}(\partial_{r}R)_{t}R^{2}dr,
\label{eq:rho-sol}
\end{eqnarray}
where $f(r)$ and $F(r)$ are arbitrary functions. 
Eq.(\ref{eq:energy-eq}) might be regarded as an energy equation 
of the dust fluid element at $r$.  From this point of view, 
the function $f(r)$ corresponds to the specific energy of the 
dust fluid element and the function $F(r)$ can be regarded
as the gravitational mass function. 
The solution is completely fixed by choosing the specific energy $f(r)$ 
and initial rest-mass density profile ${\bar \rho}(0,r)$, 
or equivalently the mass function $F(r)$. 
The shell focusing naked singularity is formed only at $r=0$.

For simplicity, hereafter we will focus on the marginally
bound case $f(r)=0$. Since our interest is on the behavior
of the space-time near the central singularity, this restriction
might not lose generality of the conclusion.
The solution of Eq.(\ref{eq:energy-eq}) is given by
\begin{equation}
R=\left({9F\over4}\right)^{1\over3}[t_{R}(r)-t]^{2\over3},
\label{eq:radius}
\end{equation}
where $t_{R}(r)$ is an arbitrary function which corresponds to the moment of 
the singularity formation. We choose the time of singularity formation as 
\begin{equation}
t_{R}(r)={r^{3/2}\over3}\sqrt{4\over F}, \label{eq:s-time}
\end{equation}
so that $R$ agrees with $r$ at $t=0$. 

In the spherically symmetric space-time, we can naturally introduce 
a ``gravitational potential'' $\Phi_{\rm G}$ defined by
\begin{equation}
\Phi_{\rm G}\equiv {F\over R}.
\end{equation}
As Hayward discussed, the gravitational potential $\Phi_{\rm G}$ 
is deeply related to the formation of trapped region and hence 
gives a measure of the strength of the gravitational field\cite{hayward}. 
To see the behavior of the gravitational potential $\Phi_{\rm G}$ and 
the velocity $(\partial_{t}R)_{r}$ of the dust fluid element, 
we adopt the following initial density configuration for $\bar \rho$
\begin{equation}
{\bar \rho}(0,r)={1\over 8\pi r^{2}}{dF\over dr}
={1\over6\pi}
\left\{1+\exp\left(-{r_{1}\over2r_{2}}\right)\right\}
\left\{1+\exp\left({{r^{2}-r_{1}^{2}\over
2r_{1}r_{2}}}\right)\right\}^{-1}, \label{eq:initial-density}
\end{equation}
where $r_{1}$ and $r_{2}$ are positive constants. 
Since a sufficient condition for the nakedness of the central shell 
focusing singularity is
$\partial_{r}^{2}\rho(0,r)<0$\cite{newman1986,jd1993}, 
the above initial density distribution necessarily leads to 
the naked singularity at the symmetric center $r=0$. 
Here note that the moment of the central shell focusing singularity 
is $t=1$ by the above initial density profile. 
The core radius of the above configuration is given by
\begin{equation}
r_{\rm core}=r_{1}+{1\over2}r_{2}.
\end{equation}
If we set appropriate $r_{1}$ and $r_{2}$, the space-time is globally
naked singular. 
In Fig.1, we depict the gravitational
potential $\Phi_{\rm G}$ with $r_{1}=2\times10^{-2}$
and $r_{2}=10^{-2}$. We see that the gravitational potential
$\Phi_{\rm G}$ is much smaller than unity
even at the moment of the central shell focusing 
naked singularity formation.
Note that form Eq.(\ref{eq:energy-eq}), $|(\partial_{t}R)_{r}|$ 
is also much smaller than unity. 
Hence in this example, the Newtonian 
approximation scheme seems to be available and in reality it is true 
as will be shown in the next section. 

\section{Newtonian Approximation}

In this section, we consider gravitational collapse 
of a spherically symmetric dust fluid in the framework of the 
Newtonian approximation and show that the Newtonian approximation 
is valid even at the moment of the central shell focusing 
naked singularity formation if the initial condition is 
appropriately set up as in the case of the example in 
the previous section. 

\subsection{Eulerian Coordinate}
 
In the Newtonian approximation, the maximal time slicing condition 
and Eulerian coordinate (for example, 
the minimal distortion gauge condition) are usually adopted.  
The line element is expressed in the following form
\begin{equation}
ds_{\rm E}^{2}=-\left(1+2\Phi_{\rm N}\right)dT^{2}+dR^{2}
+R^{2}d\Omega^{2},
\end{equation}
where $\Phi_{\rm N}$ is Newtonian gravitational potential and 
we have adopted the spherical-polar coordinate system 
as a spatial coordinates. 
The equations for the spherically symmetric dust fluid 
and Newtonian gravitational potential $\Phi_{\rm N}$
are given by
\begin{eqnarray}
\partial_{T}{\bar\rho}+{1\over R^{2}}\partial_{R}(R^{2}{\bar\rho} V)&=&0, \\
\partial_{T}V+V\partial_{R}V&=&-\partial_{R}\Phi_{\rm N},\\
{1\over R^{2}}\partial_{R}(R^{2}\partial_{R}\Phi_{\rm N})&=&4\pi{\bar\rho},
\end{eqnarray}
where $V$ is the velocity of the dust fluid element. 
The assumptions in the Newtonian approximation are 
\begin{equation}
|V|\ll1,~~~~{\rm and}~~~~
|\Phi_{\rm N}|\ll1, \label{eq:N-condition}
\end{equation}
and further
\begin{equation}
|\partial_{T}V|\ll|\partial_{R}V|,~~~~
|\partial_{T}\Phi_{\rm N}|\ll|\partial_{R}\Phi_{\rm N}|~~~~
{\rm and}~~~~
|\partial_{T}{\bar\rho}|\ll|\partial_{R}{\bar \rho}|.
\end{equation}

\subsection{Lagrangian Coordinate}

For the purpose to follow the motion of a dust sphere, 
the Lagrangian coordinate is more suitable than the Eulerian one. 
The transformation 
matrix between the Eulerian and Lagrangian coordinate systems 
is given by
\begin{eqnarray}
dT&=&d\tau, \label{eq:L-matrix1}\\
dR&=&{\dot R}d\tau+R'dx,
\label{eq:L-matrix2}
\end{eqnarray}
where regarding $\tau$ and $x$ as the independent variables,   
a dot means a partial derivative with respect to $\tau$ 
while a prime denotes a partial derivative with respect to $x$. 
Then the line element in the Lagrangian coordinate system 
is obtained as
\begin{equation}
ds_{\rm L}^{2}=-\left(1+2\Phi_{\rm N}-{\dot R}^{2}\right)d\tau^{2}
+2{\dot R}R'd\tau dx
+{R'}^{2}dx^{2}+R^{2}d\Omega^{2}.
\end{equation}
Equations for the dust fluid and Newtonian gravitational 
potential are given by
\begin{eqnarray}
F(x)&=&8\pi\int_{0}^{x}{\bar\rho}R'R^{2}dx, \\
V^{2}&=&{\dot R}^{2}=f(x)+{F(x)\over R}, \\
\Phi_{\rm N}'&=&{R'\over 2R^{2}}
F(x),\label{eq:L-poisson}
\end{eqnarray}
where $f(x)$ and $F(x)$ are regarded as 
arbitrary functions.  
Since the equation for the areal radius $R$ is the same as that 
of the LTB space-time, its solution  
for the marginally bound collapse $f(x)=0$ is given by the 
same functional form as Eq.(\ref{eq:radius}),
\begin{equation}
R=\left({9F\over4}\right)^{1\over3}[\tau_{R}(x)-\tau]^{2\over3},
\label{eq:N-radius}
\end{equation}
where $\tau_{R}(x)$ is an arbitrary function which determines 
the moment of singularity formation.

Here we consider the Newtonian approximation of the example 
given in the previous section.  
Hence we choose the moment of the singularity formation as 
\begin{equation}
\tau_{R}(x)={x^{3/2}\over 3}\sqrt{4\over F} \label{eq:N-s-time}
\end{equation}
so that $R$ agrees with $x$ at $\tau=0$. 
As for the initial density configuration, we adopt 
the same functional form as Eq.(\ref{eq:initial-density}),
\begin{equation}
{\bar \rho}(0,x)={F'\over 8\pi x^{2}}
={1\over6\pi}
\left\{1+\exp\left(-{x_{1}\over2x_{2}}\right)\right\}
\left\{1+\exp\left({{x^{2}-x_{1}^{2}\over
2x_{1}x_{2}}}\right)\right\}^{-1}, \label{eq:N-initial-density}
\end{equation}
where $x_{1}$ and $x_{2}$ are positive constants.
The above choice guarantees the regularity of all the variables 
before the singularity formation and that the central shell focusing 
singularity is formed at $\tau=1$. 

Imposing a boundary condition $\Phi_{\rm N}\rightarrow0$ for 
$x\rightarrow\infty$, the solution of Eq.(\ref{eq:L-poisson}) 
is formally expressed as
\begin{equation}
\Phi_{\rm N}
=\Phi_{\rm N1}(\tau,x)+\Phi_{\rm N2}(\tau),
\end{equation}
where
\begin{eqnarray}
\Phi_{\rm N1}(\tau,x)&\equiv&
\int_{0}^{x}{R'\over 2R^{2}}Fdx,\\
\Phi_{\rm N2}(\tau)&\equiv& -\int_{0}^{\infty}{R'
\over 2R^{2}}Fdx
=-{1\over2}\int_{0}^{\infty}{F'\over R}dx. \label{eq:Phi2-def}
\end{eqnarray}
The behavior of $F/R$ is the same as 
that of $\Phi_{\rm G}$ as shown in Fig.1. 
On the other hand, the Newtonian gravitational potential $\Phi_{\rm N}$ 
of $x_{1}=2\times10^{-2}$ and $x_{2}=10^{-2}$ 
is depicted in Fig.2(a). Form these figures, we find that the condition 
(\ref{eq:N-condition}) is satisfied even at the moment of 
the central shell focusing naked singularity formation. 
Here it is worthwhile to note that the right hand side of 
Eq.(\ref{eq:L-poisson}) at $x=0$ diverges at the
moment of the central shell focusing naked singularity formation,  
\begin{equation}
\Phi_{\rm N}'\longrightarrow {14\over 27\tau_{R(1)}^{2/3}}
~x^{-1/3}~~~~~~{\rm for}~~x\longrightarrow0~~~~{\rm at}~~\tau=1,
\end{equation}
where 
\begin{equation}
\tau_{R(1)}\equiv{1\over2}{d^{2}\tau_{R}(x)\over dx^{2}}\biggr|_{x=0}.
\end{equation}
However since the power index of $x$ is larger than $-1$, $\Phi_{\rm N}$ 
itself is finite at $x=0$ even at the moment of the central 
shell focusing singularity formation $\tau=1$. 

In order that the Newtonian approximation is successful, 
temporal derivatives of all the quantities should always 
smaller than the radial derivatives of those. Here we shall focus on 
the neighborhood of the central shell focusing naked singularity only.  
For this purpose, we introduce a new variable $w$ defined by
\begin{equation}
w\equiv \delta\tau^{-1/2}x. \label{eq:w-definition}
\end{equation}
where $\delta\tau\equiv (1-\tau)/\tau_{R(1)}$.
Then we consider a limit $\tau\rightarrow 1$ with fixed $w$. 
It should be noted that $x$ also goes to zero by this 
limiting procedure. 
The mass function $F$, rest-mass density ${\bar \rho}$ and areal radius 
$R$ behave as
\begin{eqnarray}
F &\longrightarrow&{{4\over 9}} w^{3} \delta \tau^{3/2}, \label{eq:F-ap}\\
{\bar\rho}&\longrightarrow&{1\over2\pi}\tau_{R(1)}^{-2}(3+7w^{2})^{-1}
(1+w^{2})^{-1}\delta\tau^{-2}, \label{eq:rho-ap} \\
R   &\longrightarrow&\tau_{R(1)}^{2/3}w(1+w^{2})^{2/3}\delta\tau^{7/6}. 
\label{eq:R-ap}
\end{eqnarray}
All these variables are proportional to the power of $\delta\tau$ 
and the coefficients of those are functions of $w$. 
It is easy to see that their derivatives with respect to $\tau$ or  
$x$ also take the same functional structure with respect to $\delta\tau$ and 
$w$. Thus the $x$-dependent part $\Phi_{\rm N1}$ of  
the Newtonian gravitational potential will also behave in the manner
\begin{equation}
\Phi_{\rm N1}\longrightarrow \phi_{\rm N1}(w)\delta\tau^{i},
\end{equation}
where $i$ is a constant and $\phi_{\rm N}$ is 
a function of $w$. Substituting the above equation 
into Eq.(\ref{eq:L-poisson}) and using the asymptotic behavior 
(\ref{eq:F-ap}) and (\ref{eq:R-ap}), we obtain
\begin{equation}
{d\phi_{\rm N1}(w)\over dw}\delta\tau^{i-1/2}
={2w(3+7w^{2})\over 27\tau_{R(1)}^{2/3}(1+w^{2})^{5/3}}~\delta\tau^{-1/6}.
\label{eq:Phi-ap-equation}
\end{equation}
In order that the dependence of both sides in the above
equation on $\delta\tau$ agrees with each other, $i$ should be equal to $1/3$. 
Integration of Eq.(\ref{eq:Phi-ap-equation}) leads to
\begin{equation}
\phi_{\rm N1}(w)={1\over 9\tau_{R(1)}^{2/3}(1+w^{2})^{2/3}}
\left\{7(1+w^{2})-9(1+w^{2})^{2/3}+2\right\}.
\end{equation}

In order to see the asymptotic dependence of $\Phi_{\rm N2}(\tau)$ on $\tau$, 
we differentiate Eq.(\ref{eq:Phi2-def}) to obtain
\begin{equation}
{\dot \Phi_{\rm N2}}=-{1\over2}\int_{0}^{\infty}{F'\over R^{2}}
\sqrt{F\over R}dx.\label{eq:dt-Phi}
\end{equation}
The integrand in the right hand side of the above equation 
behaves near the origin at the moment of the central shell focusing 
singularity formation as
\begin{equation}
{F'\over R^{2}}\sqrt{F\over R}\longrightarrow 
{8\over 9\tau_{R(1)}^{5/3}}x^{-7/3}~~~~~{\rm for}~~
x\longrightarrow 0~~~~~{\rm at}~~\tau=1.
\end{equation}
Therefore, the integral in Eq.(\ref{eq:dt-Phi}) 
does not have finite value at $\tau=1$. Since as shown in the above, 
this divergence comes from the irregularity of the integrand 
at the origin $x=0$, we shall 
estimate the contribution near the origin 
to the integral in Eq.(\ref{eq:dt-Phi}). We again consider the limit of 
$\delta\tau\rightarrow0$ with fixed $w$ and obtain
\begin{equation}
\int_{0}^{\infty}{F'\over R^{2}}\sqrt{F\over R}dx
\longrightarrow {8\over 9\tau_{R(1)}^{5/3}}\delta\tau^{-2/3}
\int_{0}^{\infty}{wdw \over (1+w^{2})^{5/3}}
={2\over3\tau_{R(1)}^{5/3}}\delta\tau^{-2/3}.
\end{equation}
Substituting the above equation into 
Eq.(\ref{eq:dt-Phi}) and integrating it with respect to $\tau$, 
we obtain
\begin{equation}
\Phi_{\rm N2}\longrightarrow {\delta\tau^{1/3}\over \tau_{R(1)}^{2/3}}
+\Phi_{\rm N2}(1). \label{eq:PhiN2-ap}
\end{equation}
As a result, in the limit of $\delta\tau\rightarrow0$ with 
fixed $w$, $\Phi_{\rm N}$ is expressed in the form 
\begin{equation}
\Phi_{\rm N}\longrightarrow{9+7w^{2}\over 9\tau_{R(1)}^{2/3}(1+w^{2})^{2/3}}
~\delta\tau^{1/3}+\Phi_{\rm N2}(1). \label{eq:N-Phi-ap}
\end{equation}
We depict numerically obtained $\Phi_{\rm N}$ together with the 
above asymptotic form in Fig.2(b). The asymptotic estimate agrees with the
numerical result quite well. 

Now we have known that in the limit of $\delta\tau\rightarrow 0$ with fixed
$w$, all the variables behave as
\begin{equation}
Z(\tau,x)\longrightarrow z(w)\delta\tau^{j}+{\rm constant}, \label{eq:Z-ap}
\end{equation}
where $z(w)$ is some function of $w$ and $j$ is a constant. 
The derivatives of $Z$ with respect to $T$ and $R$ are 
expressed by using its derivatives with respect to $\tau$ and $x$ as
\begin{eqnarray}
(\partial_{T}Z)_{R}&=&(\partial_{T}\tau)_{R}{\dot Z}
+(\partial_{T}x)_{R}Z', \\
(\partial_{R}Z)_{T}&=&(\partial_{R}\tau)_{T}{\dot Z}
+(\partial_{R}x)_{T}Z'.
\end{eqnarray}
From Eqs.(\ref{eq:L-matrix1}) and (\ref{eq:L-matrix2}), we find 
\begin{equation}
(\partial_{T}\tau)_{R}=1,~~(\partial_{R}\tau)_{T}=0,~~
(\partial_{T}x)_{R}=-{{\dot R}\over R'}
~~{\rm and}~~(\partial_{R}x)_{T}={1\over R'}.
\end{equation}
Then the following relation is derived,
\begin{equation}
{(\partial_{T}Z)_{R}\over (\partial_{R}Z)_{T}}
=R'{{\dot Z}\over Z'}-{\dot R}.
\end{equation}
Inserting Eq.(\ref{eq:Z-ap}) into the above equation, we obtain
\begin{equation}
{(\partial_{T}Z)_{R}\over (\partial_{R}Z)_{T}}
\longrightarrow {1\over3\tau_{R(1)}^{1/3}(1+w^{2})^{1/3}}
\left\{{1\over2}w(1+7w^{2})-jz(3+7w^{2})\left({dz\over dw}\right)^{-1}
\right\}~\delta\tau^{1\over6}.
\end{equation}
The above equation means that in the limit of $\delta\tau\rightarrow0$ 
with fixed $w$, the following inequality holds 
\begin{equation}
|(\partial_{T}Z)_{R}| \ll |(\partial_{R}Z)_{T}|.
\end{equation}
Therefore, the order counting of the Newtonian approximation 
is guaranteed even in the neighborhood of the central 
naked singularity. 

Here it is worthy to notice that in the limit of
$\delta\tau\rightarrow0$ with fixed $w$, $|{\dot Z}|$ is much larger 
than $|Z'|$ because of 
\begin{equation}
{{\dot Z}\over Z'}\longrightarrow {1\over \tau_{R(1)}}
\left({1\over2}w-jz{dw\over dz}\right)~\delta\tau^{-1/2}\longrightarrow\infty.
\end{equation}
Hence the vicinity of the central shell focusing naked singularity 
is not Newtonian situation in ordinary sense. 
In reality, the present authors have shown that 
the violent quantum particle creation occurs in such a 
situation\cite{hin2000a,hin2000b}. 
This means that the formation of the central shell focusing 
singularity is a highly relativistic phenomenon even though 
it is well described by the Newtonian approximation 
in the Eulerian coordinate system. 

\subsection{Synchronous-Comoving Coordinate}

In the LTB solution, the SC coordinate 
system is adopted. We should note that the SC coordinate system is 
different from the Lagrangian one. 
We consider the following coordinate transformation
\begin{eqnarray}
dt&=&\left\{1+\Phi_{\rm N}-{1\over2}(\partial_{\tau}R)_{x}^{2}\right\}d\tau
   -(\partial_{\tau}R)_{x}(\partial_{x}R)_{\tau}dx, \label{eq:SC-matrix1}\\
dr&=&dx. \label{eq:SC-matrix2}
\end{eqnarray}
Assuming Eq.(\ref{eq:N-condition}), derivatives of the areal 
radius $R$ with respect to 
the Lagrangian time coordinate $\tau$ and the radial coordinate $x$ 
are written as
\begin{eqnarray}
{\dot R}&=&(\partial_{\tau}t)_{x}(\partial_{t}R)_{r}
 +(\partial_{\tau}r)_{x}(\partial_{r}R)_{t}
 =\left\{1+\Phi_{\rm N}-(\partial_{\tau}R)_{x}^{2}\right\}
 (\partial_{t}R)_{x}\sim(\partial_{x}R)_{\tau},\\
R'&=&(\partial_{x}t)_{\tau}(\partial_{t}R)_{r}
+(\partial_{x}r)_{\tau}(\partial_{r}R)_{t}
=(\partial_{r}R)_{t}-(\partial_{\tau}R)_{x}(\partial_{x}R)_{\tau}
(\partial_{t}R)_{r}\sim(\partial_{r}R)_{t}. 
\end{eqnarray}
Hence the line element in the new coordinate system $t,r$ 
up to the lowest order in the sense of 
Eq.(\ref{eq:N-condition}) is written as
\begin{equation}
ds^{2}_{\rm SC}=-dt^{2}+(\partial_{r}R)_{t}^{2}dr^{2}+R^{2}d\Omega^{2}.
\end{equation}
From the above equation, we find that 
the coordinate transformation (\ref{eq:SC-matrix1}) and 
(\ref{eq:SC-matrix2}) leads to the SC coordinate system. 
Especially, the above line element is completely the same as 
the relativistic one in the marginally bound case. 
Further we obtain the equations of the lowest order  
in the SC coordinate system are given in completely the same form 
as that in the relativistic one: 
\begin{eqnarray}
(\partial_{t}R)_{r}^{2}&=&f(r)+{F(r)\over R}, \\
F(r)&=&8\pi\int_{0}^{r}{\bar \rho}(\partial_{r}R)_{t}R^{2}dr.
\end{eqnarray}
It is worthwhile to note that in the SC coordinate system, 
the Newtonian gravitational potential $\Phi_{\rm N}$ does not appear. 

\section{Basic Equations of Even Mode Perturbations}

We consider non-spherical linear perturbations in the system of the 
spherically symmetric dust ball described in the previous section.  
First, we consider perturbations in the Eulerian coordinate system. 
The line element is written as
\begin{equation}
ds_{\rm E}^{2}=-\left(1+2\Phi_{\rm N}+2\delta\Phi_{\rm N}\right)dT^{2}
+dR^{2}+R^{2}d\Omega^{2},
\end{equation}
where $\delta\Phi_{\rm N}$ is a perturbation of the Newtonian 
gravitational potential.
Using the transformation matrix (\ref{eq:L-matrix1}) 
and (\ref{eq:L-matrix2}), we obtain the perturbed line element 
in the background Lagrangian coordinate system as
\begin{equation}
ds^{2}_{\rm L}=-\left(1+2\Phi_{\rm N}+2\delta\Phi_{\rm N}
 -{\dot R}^{2}\right)d\tau^{2}+2{\dot R}R'd\tau dx
+{R'}^{2}dx^{2}+R^{2}d\Omega^{2}.
\end{equation}
Hereafter we discuss the behavior of perturbations in this coordinate
system. The density $\rho$ and 4-velocity $u^{\mu}$ 
are written in the form
\begin{eqnarray}
\rho&=&{\bar \rho}(1+\delta_{\rho}), \\
u^{\mu}&=&{\bar u}^{\mu}+\delta u^{\mu}.
\end{eqnarray}
By definition of the Lagrangian coordinate system, 
the components of the background 4-velocity is 
given by 
\begin{equation}
\left({\bar u}^{\mu}\right)=\left({\bar u}^{0},0,0,0\right).
\end{equation}
From the normalization of the 4-velocity, we find 
\begin{equation}
\delta u^{0}=-\delta\Phi_{\rm N}+{\dot R}R'\delta u^{1}.
\end{equation}

The order-counting with respect to the expansion parameter 
$\varepsilon$ of the Newtonian approximation is given by
\begin{equation}
\delta u^{0}=O(\varepsilon^{2}),~~~~\delta u^{\ell}=O(\varepsilon),
~~~~\delta_{\rho}=O(\varepsilon^{0})~~~~{\rm and}~~~~
\delta\Phi_{N}=O(\varepsilon^{2}). 
\end{equation}
Then the equations for the perturbations are given by
\begin{eqnarray}
\partial_{\tau}\delta_{\rho}
+{1\over {\bar \rho}\sqrt{\bar \gamma}}
\partial_{\ell}\left({\bar \rho}\sqrt{\bar\gamma} 
\delta u^{\ell}\right)&=&0, \label{eq:rho-eq1}\\
\partial_{\tau}\delta u_{\ell}+\partial_{\ell}\delta\Phi_{N}
&=&0, \label{eq:v-eq1}\\
{1\over \sqrt{{\bar \gamma}}}\partial_{\ell}
\left(\sqrt{\bar \gamma}{\bar \gamma}^{\ell m}\partial_{m}
\delta\Phi_{N}\right)-4\pi{\bar\rho}\delta_{\rho}&=&0,\label{eq:p-eq1}
\end{eqnarray}
where 
\begin{equation}
\sqrt{\bar \gamma}\equiv R'R^{2}\sin\theta, 
\end{equation}
and ${\bar \gamma}^{\ell m}$ is a contravariant component 
of the background 3-metric. 

Here we focus on the axisymmetric even mode of perturbations. Hence 
the perturbations are expressed in the form
\begin{eqnarray}
\delta_{\rho}&=&\sum_{l}\Delta_{\rho(l)}(\tau,x)
 P_{l}(\cos\theta), \\
\delta\Phi_{N}&=&\sum_{l}\Delta_{\Phi(l)}(\tau,x)
 P_{l}(\cos\theta), \\
\delta u_{1}&=&\sum_{l}U_{x(l)}(\tau,x)
 P_{l}(\cos\theta), \\
\delta u_{2}&=&\sum_{l}U_{\theta(l)}(\tau,x){d\over d\theta}
 P_{l}(\cos\theta), \\
\delta u_{3}&=&0.
\end{eqnarray}
From Eqs.(\ref{eq:rho-eq1}), (\ref{eq:v-eq1}) and (\ref{eq:p-eq1}), 
we obtain
\begin{eqnarray}
{\dot\Delta}_{\rho(l)}
+{1\over F'}\left({F'\over R'^{2}}U_{x(l)}
\right)'-l(l+1){U_{\theta(l)}\over R^{2}}&=&0, \label{eq:rho-eq}\\
{\dot U}_{x(l)}+\Delta_{\Phi(l)}'&=&0, \label{eq:Ux-eq}\\
{\dot U}_{\theta(l)}+\Delta_{\Phi(l)}&=&0, \label{eq:Ut-eq}\\
{1\over R'R^{2}}
\left({R^{2}\over R'}\Delta_{\Phi(l)}'\right)'-
l(l+1){\Delta_{\Phi(l)}\over R^{2}}
-4\pi{\bar\rho}\Delta_{\rho(l)}&=&0,\label{eq:P-eq}
\end{eqnarray}
Comparing the basic equations for the relativistic perturbations 
of $l=2$ in Ref.\cite{ihn1999b} to the above equations, 
we find the correspondence between the Newtonian and relativistic 
variables; $\Delta_{\Phi(2)}=-2K$, 
$U_{x(2)}=-V_{1}$ and $U_{\theta(2)}=-V_{2}$.

\section{Mass-Quadrupole Formula}

Hereafter we focus on the quadrupole mode $l=2$ and therefore 
omit the subscript $(l)$ to specify the multi-pole component 
of the perturbation variable. 
The mass-quadrupole moment $Q_{\ell m}$ is given by
\begin{equation}
Q_{\ell m}\equiv \int\rho
\left(X_{\ell}X_{m}-{1\over3}R^{2}\delta_{\ell m}\right)d^{3}X
={4\pi\over 15}Q(T){\rm diag}[-1,-1,2],
\end{equation}
where 
\begin{equation}
Q(T)\equiv\int_{0}^{\infty}{\bar\rho}\Delta_{\rho}R^{4}dR.
\label{eq:QB-relation}
\end{equation}
For a function $g(\tau,x)$ with sufficiently rapid 
fall off for $x\rightarrow \infty$, we find that
\begin{eqnarray}
{d\over dT}\int_{0}^{\infty}gdR
&=&\int_{0}^{\infty}(\partial_{T} g)_{R}dR
=\int_{0}^{\infty}\left({\dot g}-{{\dot R}\over R'}g'\right)R'dx \\
&=&\int_{0}^{\infty}\left(R'{\dot g}
 +{\dot R}'g\right)dx-\left[{\dot R}g\right]_{0}^{\infty}
=\int_{0}^{\infty}\partial_{\tau}(R'g)dx.
\end{eqnarray}
Using the above formula, we obtain
\begin{equation}
{d^{m}Q\over dT^{m}}
=\int_{0}^{\infty}
{\partial^{m}\over\partial\tau^{m}} 
\left({\bar\rho }\Delta_{\rho}R'R^{4}\right)dx
={1\over8\pi}\int_{0}^{\infty}F'
{\partial^{m}\over\partial\tau^{m}} 
\left(\Delta_{\rho}R^{2}\right)dx. \label{eq:delt-Q}
\end{equation}
The power $L_{\rm GW}$ carried by the gravitational radiation 
at the future null infinity $T+R\rightarrow\infty$ is 
given by
\begin{equation}
L_{\rm GW}={32\pi^{2}\over375}\left({d^{3}Q(u)\over du^{3}}\right)^{2},
\label{eq:G-power}
\end{equation}
where $u\equiv T-R$ is the retarded time. 
The Weyl scalar $\Psi_{4}$ carried by outgoing gravitational 
waves at the future null infinity is estimated as 
\begin{equation}
\Psi_{4}=-C_{abcd}n^{a}{\bar m}^{b}n^{c}{\bar m}^{d}
=-{3\pi\over5}{d^{4}Q(u)\over du^{4}}\sin^{2}\theta,
\label{eq:curvature}
\end{equation}
where $C_{abcd}$ is the Weyl tensor, and $n^{a}$ and ${\bar m}^{a}$ 
are two of the null tetrad basis whose components in the 
spherical polar coordinate are given by
\begin{eqnarray}
(n_{\mu})&=&{1\over \sqrt{2}}(1,-1,0,0), \\
({\bar m}_{\mu})&=&{1\over\sqrt{2}}(0,0,R,-iR\sin\theta).
\end{eqnarray}
It should be noted that the power $L_{\rm GW}$ is proportional 
to the square of the 3rd-order derivative of $Q(u)$ while 
the Weyl scalar $\Psi_{4}$ is  
proportional to the 4th-order derivatives of $Q(u)$. 

\section{Numerical Simulation}

In this section, we explain the procedure to numerically 
integrate the system of partial differential equations for 
the quadrupole perturbation variables and show the time evolution of 
the mass-quadrupole moment. 

\subsection{Basic Equation}

We assume that all the perturbation variables are regular 
before the central shell focusing singularity formation and hence 
are written in the form of the Taylor series as
\begin{eqnarray}
 \Delta_{\rho}
&=&x^{2}{\Delta}^{*}_{\rho}
 =x^{2}\sum_{n=0}^{\infty}{\Delta}^{*}_{\rho(n)}x^{2n},\\
 U_{x}
&=&x{U}^{*}_{x}=x\sum_{n=0}^{\infty}{U}^{*}_{x(n)}x^{2n}, \\
 U_{\theta}
&=&x^{2}{U}^{*}_{\theta}=x^{2}\sum_{n=0}^{\infty}
 {U}^{*}_{\theta(n)}x^{2n}, \\
\Delta_{\Phi}&=&x^{2}{\Delta}^{*}_{\Phi}=x^{2}\sum_{n=0}^{\infty}
 {\Delta}^{*}_{\Phi(n)}x^{2n}. 
\end{eqnarray}
In the numerical calculation, we focus on the above asterisked variables 
rather than the original variables. 

To obtain the solution of Eq.(\ref{eq:P-eq}), we introduce a 
variable ${\cal Q}$ defined by 
\begin{equation}
\Delta_{\Phi}={{\cal Q}(\tau,x)\over R^{3}}.
\end{equation}
Substituting the above form of $\Delta_{\Phi}$ into Eq.(\ref{eq:P-eq}), 
we obtain the equation for ${\cal Q}$ as
\begin{equation}
\left({1\over R'R^{4}}
{\cal Q}'\right)'
={F'\Delta_{\rho}\over 2R^{3}}.
\end{equation}
Integrating the above equation, we obtain
\begin{equation}
{\cal Q}'
=-{1\over2}R'R^{4}\int_{x}^{\infty}dx_{1}
{F'(x_{1})\over R^{3}(\tau,x_{1})}
\Delta_{\rho}(\tau,x_{1}),
\end{equation}
where we have chosen the integration constant so that 
${\cal Q}'$ is finite for $x\rightarrow\infty$. 
Further integration leads to
\begin{eqnarray}
{\cal Q}(\tau,x)&=&-{1\over2}\int_{0}^{x}dx_{1}R'(\tau,x_{1})
 R^{4}(\tau,x_{1})\int_{x_{1}}^{\infty}dx_{2}
 {F'(x_{2})\over R^{3}(\tau,x_{2})}
 \Delta_{\rho}(\tau,x_{2}) \nonumber \\
&=&-{1\over2}\int_{0}^{x}dx_{1}R'(\tau,x_{1})
 R^{4}(\tau,x_{1})\int_{x}^{\infty}dx_{2}
 {F'(x_{2})\over R^{3}(\tau,x_{2})}
 \Delta_{\rho}(\tau,x_{2}) \nonumber \\
& &-{1\over2}\int_{0}^{x}dx_{2}{F'(x_{2})\over R^{3}(\tau,x_{2})}
 \Delta_{\rho}(\tau,x_{2})
 \int_{0}^{x_{2}}dx_{1}R'(\tau,x_{1})R^{4}(\tau,x_{1}) \nonumber \\
&=&-{1\over10}\left\{R^{5}{\cal A}(\tau,x)+x^{7}{\cal B}(\tau,x)\right\}.
\end{eqnarray}
where 
\begin{eqnarray}
{\cal A}(\tau,x)&\equiv& \int_{x}^{\infty}dx_{1}
 {F'(x_{1})\over R^{3}(\tau,x_{1})}
 \Delta_{\rho}(\tau,x_{1}), \label{eq:cal-A}\\
{\cal B}(\tau,x)&\equiv&{1\over x^{7}}\int_{0}^{x}dx_{1}
  F'(x_{1})\Delta_{\rho}(\tau,x_{1})R^{2}(\tau,x_{1}), \label{eq:cal-B}
\end{eqnarray}
and we have chosen the integration constant so that 
${\cal Q}$ vanishes at the origin, $x=0$. 
Hence ${\Delta}^{*}_{\Phi}$ is written as
\begin{equation}
{\Delta}_{\Phi}^{*}=-{1\over10}
\left\{\left({R\over x}\right)^{2}{\cal A}
+x^{2}\left({x\over R}\right)^{3}{\cal B}\right\}. \label{eq:Dphi-eq2}
\end{equation}

For the numerical calculation, we rewrite the 
basic equations (\ref{eq:rho-eq})$-$(\ref{eq:Ut-eq}) 
for the perturbation variables. Differentiating the conservation 
law (\ref{eq:rho-eq}) with respect to 
time $\tau$ and then using Eqs.(\ref{eq:rho-eq})$-$(\ref{eq:Ut-eq}), 
we obtain a second order differential equation with respect 
to $\tau$ for ${\Delta}^{*}_{\rho}$ as
\begin{eqnarray}
{\partial {V}^{*}_{\rho}\over \partial\tau}&=&
-2{{\dot R}'\over R'}{V}^{*}_{\rho}+{F'\over 2R'R^{2}}
{\Delta}^{*}_{\rho} 
+{2\over x{R'}^{2}}
\left({{\dot R}'\over R'}\right)'{U}^{*}_{x}
-{12\over R^{2}}\left({{\dot R}\over R}-{{\dot R}'\over R'}\right)
{U}^{*}_{\theta}
+{R^{2}\over xF'R'}
\left({F'\over R'R^{2}}\right)'{D}^{*}_{\Phi}, \label{eq:Vrho-eq2}\\
{\partial {\Delta}^{*}_{\rho}\over \partial\tau}&=&{V}^{*}_{\rho},
\label{eq:Drho-eq2}\\
\end{eqnarray}
where 
\begin{equation}
{D}^{*}_{\Phi}\equiv {1\over x}(x^{2}{\Delta}^{*}_{\Phi})'
=-{R' \over 10}
\left\{2\left({R\over x}\right){\cal A}
-3x^{2}\left({x\over R}\right)^{4}{\cal B}\right\}. \label{eq:Dphi-def}
\end{equation}
The equations for ${U}^{*}_{x}$ and ${U}^{*}_{\theta}$ are 
written as
\begin{eqnarray}
{\partial{U}^{*}_{x}\over \partial\tau}&=&-{D}^{*}_{\Phi}, 
\label{eq:Ux-eq2}\\
{\partial{U}^{*}_{\theta}\over\partial\tau}&=&-{\Delta}^{*}_{\Phi}.
\label{eq:Ut-eq2}
\end{eqnarray}
Eqs.(\ref{eq:Dphi-eq2})$-$(\ref{eq:Ut-eq2}) constitute a closed system 
of equations, which is solved numerically. 

\subsection{Background and Initial Condition}

The background solution is given by (\ref{eq:N-radius}), 
(\ref{eq:N-s-time}) and (\ref{eq:N-initial-density}).
The parameters $x_{1}$ and $x_{2}$ in ${\bar\rho}(0,x)$ 
are chosen to be $x_{1}=2\times10^{-2}$  
and $x_{2}=10^{-2}$, respectively. Then the core radius $x_{\rm core}$ 
of the background rest-mass density distribution becomes
\begin{equation}
x_{\rm core}=x_{1}+{1\over2}x_{2}=1.25\times10^{-2}.
\end{equation}

We start the numerical integration at $\tau=0$. 
The initial data of the density perturbation 
${\Delta}^{*}_{\rho}$ is set up in the form
\begin{equation}
{\Delta}^{*}_{\rho}(0,x)=\exp\left\{-{1\over2}
\left({x\over \sigma_{\rho} x_{\rm c}}\right)^{2}
\right\}. \label{eq:D-eq2}
\end{equation}
where $\sigma_{\rho}$ is a positive constant smaller than unity. 
We set $\sigma_{\rho}=0.5$. 

From Eqs.(\ref{eq:Ux-eq}) and (\ref{eq:Ut-eq}), we find
\begin{equation}
\partial_{\tau}\left(U_{x}-U_{\theta}'\right)=0.
\end{equation}
Hence $U_{x}$ is written in the form
\begin{equation}
U_{x}=U_{\theta}'+C_{x}(x),
\end{equation}
where $C_{x}$ is an arbitrary function. 
As for the initial data, we set $U_{\theta}=0$ and 
\begin{equation}
C_{x}=c_{x}x^{3}\exp\left\{
{-{1\over2}\left({x\over \sigma_{x}x_{\rm c}}\right)^{2}}\right\},
\end{equation}
where $\sigma_{x}$ is a positive constant smaller than unity and 
$c_{x}$ is also a constant. 
From Eq.(\ref{eq:rho-eq}), the initial data for ${V}^{*}_{\rho}$ 
should be 
\begin{equation}
{V}^{*}_{\rho}(0,x)=-{1\over x^{2}F'}(F'C_{x})'.
\end{equation}

\subsection{Numerical Scheme}

The temporal integration of Eqs.(\ref{eq:Vrho-eq2})$-$(\ref{eq:Ut-eq2}) 
is performed by the second order Runge-Kutta method. 
In order to obtain ${\Delta}^{*}_{\Phi}$ and ${D}^{*}_{\Phi}$, 
we numerically integrate Eqs.(\ref{eq:cal-A}) and 
(\ref{eq:cal-B}) also by the second order Runge-Kutta method 
and then substitute the results into 
Eqs.(\ref{eq:Dphi-eq2}) and (\ref{eq:Dphi-def}). 

The number of spatial grid points is $5\times 10^{4}$ and the  
outermost grid point corresponds to $x=10x_{\rm c}$. 
Since we numerically integrate the second order differential equation 
with respect to $\tau$ for the density perturbation ${\Delta}^{*}_{\rho}$, 
the conservation law (\ref{eq:rho-eq}) is not trivially satisfied 
by the numerical error. Hence we use Eq.(\ref{eq:rho-eq}) as 
a check of accuracy of the numerical integration. We define the 
error function $\cal E$ as
\begin{equation}
{\cal E}\equiv 4\biggl|{V}^{*}_{\rho}
+{{{U}^{*}_{x}}'\over x R'^{2}}
+{{U}^{*}_{x}\over x^{2}F'}
\left({xF'\over R'^{2}}\right)'
-{6\over R^{2}}{U}^{*}_{\theta}\biggr|
\left[|{V}^{*}_{\rho}|
+\biggl|{{{U}^{*}_{x}}'\over x R'^{2}}\biggr|
+\biggl|{{U}^{*}_{x}\over x^{2}F'}
\left({xF'\over R'^{2}}\right)'\biggr|
+\biggl|{6\over R^{2}}{U}^{*}_{\theta}\biggr|\right]^{-1}.
\end{equation}
We perform the numerical integration until $1-\tau=2\times10^{-4}$. 
The function ${\cal E}$ in $x\leq x_{\rm c}$ is less than $10^{-3}$.  
Hence the numerical integration is almost consistently performed 
near the origin and hence we can observe the asymptotic behavior 
of the contribution near the central shell focusing singularity to 
the mass-quadrupole moment in the limit of  $\tau\rightarrow1$. 

\subsection{Behavior of the Mass-Quadrupole Moment}

In Fig3, we show $d^{2}Q/dT^{2}$, $d^{3}Q/dT^{3}$ 
and $d^{4}Q/dT^{4}$. The numerical integration was performed 
from $1-\tau=1$ to $1-\tau=2\times10^{-4}$. 
From this figure, we find that those variables 
do not show power-law behavior with respect to $1-\tau$ but 
rather their behavior is oscillatory and growing. 
At this stage, the mass-quadrupole moment is determined mainly by the 
motion of dust fluid in the outside region and the contribution 
in the neighborhood of the central shell focusing singularity is 
still negligible. However, the contribution near the origin will be  
dominant in the limit $\tau\rightarrow1$ as will be shown in 
the next section.

\section{Asymptotic Analysis of the Perturbations}

In order to get information of the asymptotic behavior of the 
mass-quadrupole moment, we should carefully examine the asymptotic behavior 
of the perturbation variables near the origin. 
For this purpose, we introduce $w$ defined in Eq.(\ref{eq:w-definition}) 
and then consider the limit $\tau\rightarrow 1$ with fixed $w$. 
Since all the background variables appearing in the 
equations of the perturbations are proportional to the power 
of $\delta\tau$ and the coefficients of those are the functions of $w$ 
as Eqs.(\ref{eq:F-ap})$-$(\ref{eq:R-ap}) and (\ref{eq:N-Phi-ap}),  
we expect that the perturbation variables 
also behave in the same manner as the background variables and 
hence we assume 
\begin{eqnarray}
{\Delta}^{*}_{\rho}&=&{\delta}^{*}_{\rho}(w)\delta\tau^{-p}, 
\label{eq:rho-asp}\\
{U}^{*}_{x}&=&{1\over x}\partial_{x}(x^{2}{U}^{*}_{\theta})
=\tau_{R(1)}^{1/3}
\left(w{d{u}^{*}_{\theta}\over dw}+2{u}^{*}_{\theta}\right)
\delta\tau^{-q}, \\
{U}^{*}_{\theta}&=&\tau_{R(1)}^{1/3}{u}^{*}_{\theta}(w)\delta\tau^{-q}, 
\label{eq:Ut-asp}\\
{\Delta}^{*}_{\Phi}&=&\tau_{R(1)}^{-2/3}
{\delta}^{*}_{\Phi}(w)\delta\tau^{-r},\label{eq:P-asp}
\end{eqnarray}
where we have assumed that the contribution of $C_{x}$ to 
${U}^{*}_{x}$ is negligible in the limit of $\tau\rightarrow1$. 
In Fig.4, we depict the numerical results for the asterisked variables at
the origin as functions of $1-\tau$ in the case of $C_{x}=0$. 
From this figure, we can easily find that the hatted variables show the 
power-low behavior and grow monotonically. The power indices are
estimated as
\begin{equation}
p=1.70,~~~~q=0.38~~~~{\rm and}~~~~r=1.37. \label{eq:N-results}
\end{equation}
Due to this growing behavior, in reality, the contribution of $C_{x}$ to 
${U}^{*}_{x}$ becomes negligible in the limit $\tau\rightarrow1$. 
Although we do not show the case of $C_{x}\neq0$, we have confirmed 
that the behavior of the asterisked variables in the case of $C_{x}\neq0$ 
is almost identical to the case of $C_{x}=0$ in the limit 
$\tau\rightarrow1$.
 
Next we depict the following normalized perturbation variables 
with respect to $w$ for $x<0.05x_{\rm core}$ 
at various time steps for the time interval 
$2.0\times10^{-4}\leq 1-\tau\leq6.6\times10^{-3}$ in Figs.5, 6 and 7;
\begin{eqnarray}
{\Delta}^{\dagger}_{\rho}&\equiv& 
{\Delta}^{*}_{\rho}(\tau,x)/{\Delta}^{*}_{\rho}(\tau,0), \\
{U}^{\dagger}_{\theta}&\equiv& 
{U}^{*}_{\theta}(\tau,x)/{U}^{*}_{\theta}(\tau,0), \\
{\Delta}^{\dagger}_{\Phi}&\equiv& 
{\Delta}^{*}_{\Phi}(\tau,x)/{\Delta}^{*}_{\Phi}(\tau,0).
\end{eqnarray}
Note that $\Delta_{\rho}^{*}(\tau,0)$ at $1-\tau=2\times10^{-4}$ 
becomes 381 times larger than that at $1-\tau=6.6\times10^{-3}$. 
As expected, the spatial configuration of these variables 
with respect to $w$ are almost 
time independent. Hence the 
assumptions (\ref{eq:rho-asp})$-$(\ref{eq:P-asp}) are justified 
by the numerical calculation. 

Now, by virtue of our knowledge about the asymptotic forms 
(\ref{eq:rho-asp})$-$(\ref{eq:P-asp}), more rigorous 
discussion about the evolution of the mass-quadrupole moment 
is possible. Substituting 
Eqs.(\ref{eq:rho-asp})$-$(\ref{eq:P-asp}) into
Eqs.(\ref{eq:rho-eq})$-$(\ref{eq:P-eq}), 
and using the asymptotic behavior of the 
background variables (\ref{eq:F-ap})$-$(\ref{eq:R-ap}), 
we obtain 
\begin{equation}
\left(w{d{\delta}^{*}_{\rho}\over dw}
+2p{\delta}^{*}_{\rho}\right)\delta\tau^{-p-1}
+\left[{18\over w^{4}}{d\over dw}\left\{{w^{2}(1+w^{2})^{2/3}
\over (3+7w^{2})^{2}}{d\over dw}(w^{2}{u}^{*}_{\theta})\right\}
-{12{u}^{*}_{\theta}\over w^{2}(1+w^{2})^{4/3}}\right]
\delta\tau^{-q-7/3}=0,\label{eq:conti-ap}
\end{equation}
\begin{equation}
\left(w{d{u}^{*}_{\theta}\over dw}+2q{u}^{*}_{\theta}\right)
\delta\tau^{-q-1}+2{\delta}^{*}_{\Phi}\delta\tau^{-r}
=0, \label{Euler-ap}
\end{equation}
and
\begin{equation}
\left[{d\over dw}\left\{{w^{2}(1+w^{2})^{5/3}\over 3+7w^{2}}
{d\over dw}(w^{2}{\delta}^{*}_{\Phi})\right\}
-{2w^{2}(3+7w^{2})\over3(1+w^{2})^{1/3}}{\delta}^{*}_{\Phi}\right]
\delta\tau^{-r+5/3}
-{2\over9}w^{4}{\delta}^{*}_{\rho}\delta\tau^{-p+2}=0.\label{eq:P-ap}
\end{equation}
Since the power of $\delta\tau$ should be balanced 
in each equation, we obtain
\begin{equation}
q=p-{4\over3}~~~~~~{\rm and}~~~~~~r=p-{1\over3}. \label{eq:power}
\end{equation}
Eqs.(\ref{eq:conti-ap})$-$(\ref{eq:P-ap}) constitute a closed system 
of ordinary differential equations. 
In order to obtain a solution of the above equations, 
we also need numerical integration. However, here we are going to search for 
regular and gentle solutions of the ordinary differential equations. 
This is much easier than the previous numerical integration of 
the partial differential equations of which solutions 
show singular behavior. Further high numerical accuracy 
is guaranteed and therefore the following analysis might be 
called ``semi-analytic''.  

By an appropriate manipulation\cite{math}, we obtain a single decoupled 
equation for ${u}^{*}_{\theta}$ as
\begin{equation}
{d^{4}{u}^{*}_{\theta}\over dy^{4}}
+c_{3}{d^{3}{u}^{*}_{\theta}\over dy^{3}}
+c_{2}{d^{2}{u}^{*}_{\theta}\over dy^{2}}
+c_{1}{d{u}^{*}_{\theta}\over dy}
+c_{0}{u}^{*}_{\theta}=0, \label{eq:u-eq}
\end{equation}
where $y\equiv w^2$ and 
\begin{eqnarray}
c_{0}&=&-2\{-24(9 + 26y + 21y^2) \nonumber \\
&+&3q^2(63 + 414y + 1016y^2 + 1106y^3 + 441y^4) \nonumber \\
&+& q(252 + 1323y + 3149y^2 + 3465y^3 + 1323y^4)\} \nonumber \\
&\times& \left\{9y^3(1 + y)^3 (3 + 7y)^2\right\}^{-1}, \\
&& \nonumber \\
c_{1}&=&\{378 + 2196y + 4758y^2 + 2006y^3 - 4424y^4 - 3234y^5 
\nonumber \\
&+&3q^2(1 + y)^2 (189 + 1035y + 1911y^2 + 1225y^3) \nonumber \\
&+& q(1323 + 7893y + 18966y^2 + 21202y^3 + 9247y^4 + 
    441y^5)\} \nonumber \\
&\times&\{18y^3(1 + y)^3(3 + 7y)^2\}^{-1}, \\
&& \nonumber \\
c_{2}&=&\{1269 + 7731y + 18453y^2 + 19565y^3 + 7350y^4 
+9q^2(3 + 10y + 7y^2)^2 \nonumber \\
&+& 6q(153 + 993y + 2387y^2 + 2527y^3 + 980y^4)\} \nonumber \\
&\times&\{9y^2(1 + y)^2(3 + 7y)^2\}^{-1}, \\
&& \nonumber \\
c_{3}&=&\{159 + 506y + 427y^2 
+ 12q(3 + 10y + 7y^2)\} \nonumber \\
&\times& \{6y(1 + y)(3 + 7y)\}^{-1}.
\end{eqnarray}
Giving an appropriate boundary condition, we can numerically solve 
Eqs.(\ref{eq:conti-ap})$-$(\ref{eq:P-ap}) as a kind of the eigen 
value problem to obtain $q$ and the solution for 
${u}^{*}_{\theta}$. 

The boundary condition at $y=0$ is given by the following procedure. 
First, we assume that ${\delta}^{*}_{\rho}$, ${u}^{*}_{\theta}$ 
and ${\delta}^{*}_{\Phi}$ are $C^{\infty}$ functions. 
By this assumption, the perturbation variables are written 
in the form of the Taylor series as, 
\begin{eqnarray}
{\delta^{*}_{\rho}}&=&\sum_{m=0}^{\infty}\delta_{(m)}y^{m}, 
\label{eq:p-rho-def} \\
{u^{*}_{\theta}}&=&\sum_{m=0}^{\infty}u_{(m)}y^{m}, \\
{\delta^{*}_{\Phi}}&=&\sum_{m=0}^{\infty}\Delta_{(m)}y^{m}. 
\label{eq:p-Phi-def}
\end{eqnarray}
Then substituting the above expressions into 
Eqs.(\ref{eq:rho-eq})$-$(\ref{eq:P-eq}), we obtain 
\begin{eqnarray}
\delta_{(0)}\left({4\over3} + q\right) - 32u_{(0)} + 14u_{(1)}&=&0, \\
\delta_{(1)}\left({7\over3}+ q\right) + {1568\over9}u_{(0)}
 - 104u_{(1)} + 36u_{(2)}&=&0,\\
\delta_{(2)}\left({10\over3} + q\right) 
- {18848\over7}u_{(0)} + {1388\over3}u_{(1)} - 208u_{(2)} 
+  66u_{(3)}&=&0, \\
&&\nonumber \\
(q+m)u_{(m)}+\Delta_{(m)}&=&0, \\
&&\nonumber \\
-28\Delta_{(0)}+ 21\Delta_{(1)} - \delta_{(0)}&=&0, \\
148\Delta_{(0)}- 3(46\Delta_{(1)}-54\Delta_{(2)}+\delta_{(1)})&=&0, \\
-{620\over81}\Delta_{(0)} + {119\over18}\Delta_{(1)}
- 4\Delta_{(2)} + {11\over2}\Delta_{(3)} - 
 {1\over18}\delta_{(2)}&=&0,
\end{eqnarray}
where $m$ is non-negative integer. 
From the above equations, we obtain 
\begin{eqnarray}
u_{(1)}&=&{4(-24 + 28q + 21q^2)\over 21(2 + 7q + 3q^2)}u_{(0)}, \\
u_{(2)}&=&{2(5248 + 4822q + 3765q^2 + 2646q^3 + 567q^4)\over
 567(2 + 7q + 3q^2)(12 + 13q + 3q^2)}u_{(0)},\\
u_{(3)}&=&-{4(2174336 + 3606320q + 2343504q^2 + 889242q^3 + 
   272511q^4 + 62370q^5 + 6237q^6)\over
 18711(2 + 7q + 3q^2)(12 + 13q + 3q^2)
  (28 + 19q + 3q^2)}u_{(0)}.
\end{eqnarray}
Since Eq.(\ref{eq:u-eq}) is linear, the value 
of $u_{(0)}$ has no special meaning and hence 
we set $u_{(0)}=1$, for simplicity. 
Then the boundary condition at $y=0$ for Eq.(\ref{eq:u-eq}) is 
uniquely determined as
\begin{equation}
{u}^{*}_{\theta}|_{y=0}=u_{(0)}=1,~~~~
{d{u}^{*}_{\theta}\over dy}\biggr|_{y=0}=u_{(1)},~~~~
{d^{2}{u}^{*}_{\theta}\over dy^{2}}\biggr|_{y=0}=2u_{(2)},~~~~
{\rm and}~~~~
{d^{3}{u}^{*}_{\theta}\over dy^{3}}\biggr|_{y=0}=6u_{(3)}. 
\end{equation}
We numerically integrate 
Eq.(\ref{eq:u-eq}) outward from $y=0$ by 4th-order Runge-Kutta method. 
The behavior of ${u}^{*}_{\theta}$ depends on the value 
of $q$. 

In order to set up the boundary condition at the outer numerical 
boundary, we consider the behavior of Eq.(\ref{eq:u-eq}) in the 
limit of $y\rightarrow \infty$, 
\begin{equation}
{d^{4}u_{\theta}^{*}\over dy^{4}}
+{1\over6y}(12q+61){d^{3}u_{\theta}^{*}\over dy^{3}}
+{1\over3y^{2}}(3q^{2}+40q+50){d^{2}u_{\theta}^{*}\over dy^{2}}
+{1\over6y^{3}}(25q^{2}+3q-22){du_{\theta}^{*}\over dy}
-{6\over y^{4}}q(q+1)u_{\theta}^{*}=0. \label{eq:u-eq-ap}
\end{equation}
Therefore, $u_{\theta}^{*}$ behaves as 
\begin{equation}
{u}^{*}_{\theta}\longrightarrow {\rm Const.}\times y^{k}
~~~~~~{\rm for}~~~~~~y\longrightarrow\infty.
\end{equation}
Inserting the above equation into Eq.(\ref{eq:u-eq-ap}), we obtain a 
4th-order algebraic equation for $k$. The solutions of 
this equation are given by
\begin{equation}
k=-q,~-(q+1),~-{9\over2}~~{\rm and}~~{4\over3}
\end{equation}
In order to pick up a solution from the above four independent 
asymptotic solutions, we consider $U_{\theta}^{*}$ in the 
limit of $y\rightarrow\infty$, 
\begin{equation}
U_{\theta}^{*}\longrightarrow {\rm Const.}\times y^{k}\delta\tau^{-q}
={\rm Const.}\times x^{2k}\delta\tau^{-(k+q)}.
\end{equation}
Here we impose a condition that $U_{\theta}^{*}$ is non-zero and finite for 
$0<x<\epsilon$ at $\delta\tau=0$, 
where $\epsilon$ is a positive infinitesimal number. 
This condition leads to
\begin{equation}
k=-q.
\end{equation}
Hence by varying $q$, we search for the solution which behaves  
\begin{equation}
{u}^{*}_{\theta}\longrightarrow {\rm Const.}\times y^{-q}
~~~~~~{\rm for}~~~~~~y\longrightarrow\infty. 
\end{equation}
The numerical calculation reveals that the above behavior is realized when 
\begin{equation}
q=0.3672~.
\end{equation}
From Eq.(\ref{eq:power}), we obtain
\begin{equation}
p=1.701~~~~~{\rm and}~~~~~~r=1.367~. \label{eq:ap-results}
\end{equation}
The above values agree with the numerical results (\ref{eq:N-results}) 
quite well. We depict the data for ${\delta}^{*}_{\rho}/\delta_{(0)}$, 
${u}^{*}_{\theta}$ and ${\delta}^{*}_{\Phi}/\Delta_{(0)}$ 
in Figs.5, 6 and 7 together with the variables ${\Delta}^{\dagger}_{\rho}$, 
${U}^{\dagger}_{\theta}$ and ${\Delta}^{\dagger}_{\Phi}$ 
obtained by the numerical calculation 
of Eqs.(\ref{eq:Vrho-eq2})$-$(\ref{eq:Ut-eq2}), 
where $\delta_{(0)}$ and $\Delta_{(0)}$ are defined by 
Eqs.(\ref{eq:p-rho-def}) and (\ref{eq:p-Phi-def}), respectively. 
We see the quite excellent agreement. 

Now we examine the mass-quadrupole moment $Q(T)$ and its time-derivatives 
$d^{m}Q/dT^{m}$. 
In order to see the contribution of the central singularity 
to $d^{m}Q/dT^{m}$, we consider the integrand in the right hand side of 
Eq.(\ref{eq:delt-Q}). 
Using Eqs.(\ref{eq:F-ap}), (\ref{eq:R-ap}) and (\ref{eq:rho-asp}), 
we obtain
\begin{eqnarray}
F'\Delta_{\rho}R^2&\longrightarrow&
{4\over3}\tau_{R(1)}^{4/3}w^{6}(1+w^{2})^{4/3}{\delta}^{*}_{\rho}(w) 
~\delta\tau^{13/3-p}\nonumber \\
&=&{4\over3}\tau_{R(1)}^{4/3}x^{2(13/3-p)}w^{2(p-4/3)}
(1+w^{2})^{4/3}{\delta}^{*}_{\rho}(w).
\end{eqnarray}
From the above equation, we obtain
\begin{eqnarray} 
I^{(m)}(\tau,x)&\equiv&
{\partial^{m}\over\partial\tau^{m}}\left(F'\Delta_{\rho}R^{2}\right)
\longrightarrow
{2^{2-m}\over3}\tau_{R(1)}^{4/3-m}
\delta\tau^{13/3-p-m}w^{26/3-2p-2m} \nonumber \\
&\times&\left(w^{3}{d\over dw}\right)^{m}
\left\{w^{2(p-4/3)}(1+w^{2})^{4/3}{\delta}^{*}_{\rho}(w)\right\}.
\label{eq:Im-def}
\end{eqnarray}

We consider the integral of $I^{(m)}$ from $x=0$ to $x=x_{\rm o}$ 
to see the contribution of the central shell focusing naked singularity to 
the time derivatives of the mass-quadrupole moment. 
Here we take a limit $\delta\tau\rightarrow0$ with 
fixed $w_{o}\equiv x_{\rm o}\delta\tau^{-1/2}$ and 
then consider the limit $w_{o}\rightarrow\infty$. 
As a result, we obtain
\begin{eqnarray}
\int_{0}^{x_{\rm o}}I^{(m)}(\tau,x)dx
&=&\delta\tau^{1/2}\int_{0}^{w_{o}}
I^{(m)}(\tau,\delta\tau^{1/2} w)dw  \nonumber \\
&\longrightarrow&{2^{2-m}\over3}\tau_{R(1)}^{4/3-m}
\delta\tau^{29/6-p-m} \nonumber \\
&\times&\int_{0}^{\infty}
w^{26/3-2p-2m}
\left(w^{3}{d\over dw}\right)^{m}
\left\{w^{2(p-4/3)}(1+w^{2})^{4/3}{\delta}^{*}_{\rho}(w)\right\}dw.
\end{eqnarray}
The above equation and Eq.(\ref{eq:ap-results}) 
show that the contribution of the 
central singularity to $d^{m}Q/dT^{m}$ diverges for $\tau\rightarrow1$ 
if and only if $m$ is larger than or equal to 4. 
This result and the quadrupole formula imply that the metric perturbation 
corresponding to the gravitational radiation and its first order 
temporal derivative is finite but the second order temporal derivative 
diverges. Hence the power $L_{\rm GW}$ of the gravitational radiation 
is finite but the curvature $\Psi_{4}$ carried by the gravitational waves 
from the central naked singularity diverges; see Eqs.(\ref{eq:G-power}) 
and (\ref{eq:curvature}). This agrees with 
the relativistic perturbation analysis. Further, we find that 
in the limit of $\tau\rightarrow1$,
\begin{equation}
\Psi_{4}=-{3\pi\over5}{d^{4}Q\over dT^{4}}
\propto\delta\tau^{5/6-p}\propto(1-\tau)^{-0.867}.
\end{equation}
This result is also consistent with the relativistic 
perturbation analysis\cite{ihn1999b}.

\section{Summary and Concluding Remarks}

We analyzed the even mode perturbations of $l=2$ in the 
spherically symmetric dust collapse in the framework 
of the Newtonian approximation and estimated the gravitational 
radiation generated by these perturbations by the quadrupole 
formula. 
Since we treat separately the dynamics of the matter perturbations 
and the gravitational 
waves in the wave zone, we can estimate 
the asymptotic behavior semi-analytically, where 
``semi-analytically'' means that we know it by solving 
the gentle ordinary differential equations.
This is the great advantage of the Newtonian approximation. 

As a result, we found that the power carried 
by the gravitational waves from the neighborhood of the 
naked singularity at the symmetric center is finite. 
However, the space-time curvature 
associated to the gravitational waves becomes infinite 
in accordance with the power law. 
This result is consistent with recently performed 
relativistic perturbation analysis by Iguchi et al.\cite{ihn1999b}. 
Furthermore, the power index obtained by the Newtonian analysis 
also agrees with the relativistic perturbation analysis 
quite well. 

The agreement between the results of the Newtonian and 
relativistic analyses implies that the perturbations 
themselves are always confined within the range to 
which the Newtonian approximation is applicable. 
Here we will focus on the metric perturbation, ${\Delta}^{*}_{\Phi}$. 
Since the asymptotic solution of ${\Delta}^{*}_{\Phi}$ has the same form 
as Eq.(\ref{eq:Z-ap}), we immediately find that 
in the limit of $\delta\tau\rightarrow0$ with fixed $w$, 
\begin{equation}
{\partial_{T}{\Delta}^{*}_{\Phi}\over \partial_{R}{\Delta}^{*}_{\Phi}}\propto 
\delta\tau^{1/6},
\end{equation}
and hence the assumption 
$|\partial_{T}{\Delta}^{*}_{\Phi}|\ll|\partial_{R}{\Delta}^{*}_{\Phi}|$
of the Newtonian approximation is valid in the Eulerian coordinate
system. We can also see 2nd-order derivatives.
In the same limit, we find 
\begin{equation}
{\partial_{T}\partial_{R}{\Delta}^{*}_{\Phi}
\over \partial_{R}^{2}{\Delta}^{*}_{\Phi}}\propto
\delta\tau^{1/6}~~~~{\rm and}~~~~
{\partial_{T}^{2}{\Delta}^{*}_{\Phi}
\over \partial_{R}\partial_{T}{\Delta}^{*}_{\Phi}}\propto 
\delta\tau^{1/6}.
\end{equation}
From the above equations, we obtain 
\begin{equation}
{\partial_{T}^{2}{\Delta}^{*}_{\Phi}
\over \partial_{R}^{2}{\Delta}^{*}_{\Phi}}\propto 
\delta\tau^{1/3}.
\end{equation}
The above equation means that in the limit of $\delta\tau\rightarrow0$ 
with fixed $w$, the following inequality is also satisfied,
\begin{equation}
|\partial_{T}^{2}{\Delta}^{*}_{\Phi}|\ll
|\partial_{R}^{2}{\Delta}^{*}_{\Phi}|.
\end{equation}
The above inequality implies 
that the wave equation for the metric perturbation 
$\Delta_{\Phi}$ is well approximated by the Poisson-type equation 
if we adopt the Eulerian coordinate system. 

However as mentioned in Sec.2, the gravitational collapse producing 
the shell focusing globally naked singularity is not Newtonian 
in ordinary sense. The same is true for the perturbation variables because 
$|{\dot\Delta_{\Phi}}/\Delta_{\Phi}'|\gg1$ 
in the limit of $\delta\tau\rightarrow0$ with fixed $w$.  
Even though the Newtonian approximation is valid for the 
Eulerian coordinate system, 
the Newtonian order counting breaks down if we adopt the
Lagrangian coordinate system as the spatial coordinates.

\vskip0.5cm
{\large{\bf Acknowledgements}}

We are grateful to H.~Sato for his continuous encouragement.
We are also grateful to T.~Nakamura, H.~Kodama, T.P.~Singh, 
A.~Ishibashi and S.S.~Deshingkar for helpful discussions.
This work was supported by the 
Grant-in-Aid for Scientific Research (No.~05540)
and for Creative Basic Research (No.~09NP0801)
from the Japanese Ministry of
Education, Science, Sports and Culture.

\newpage
\begin{figure}
        \centerline{\epsfxsize 15cm \epsfysize 10cm \epsfbox{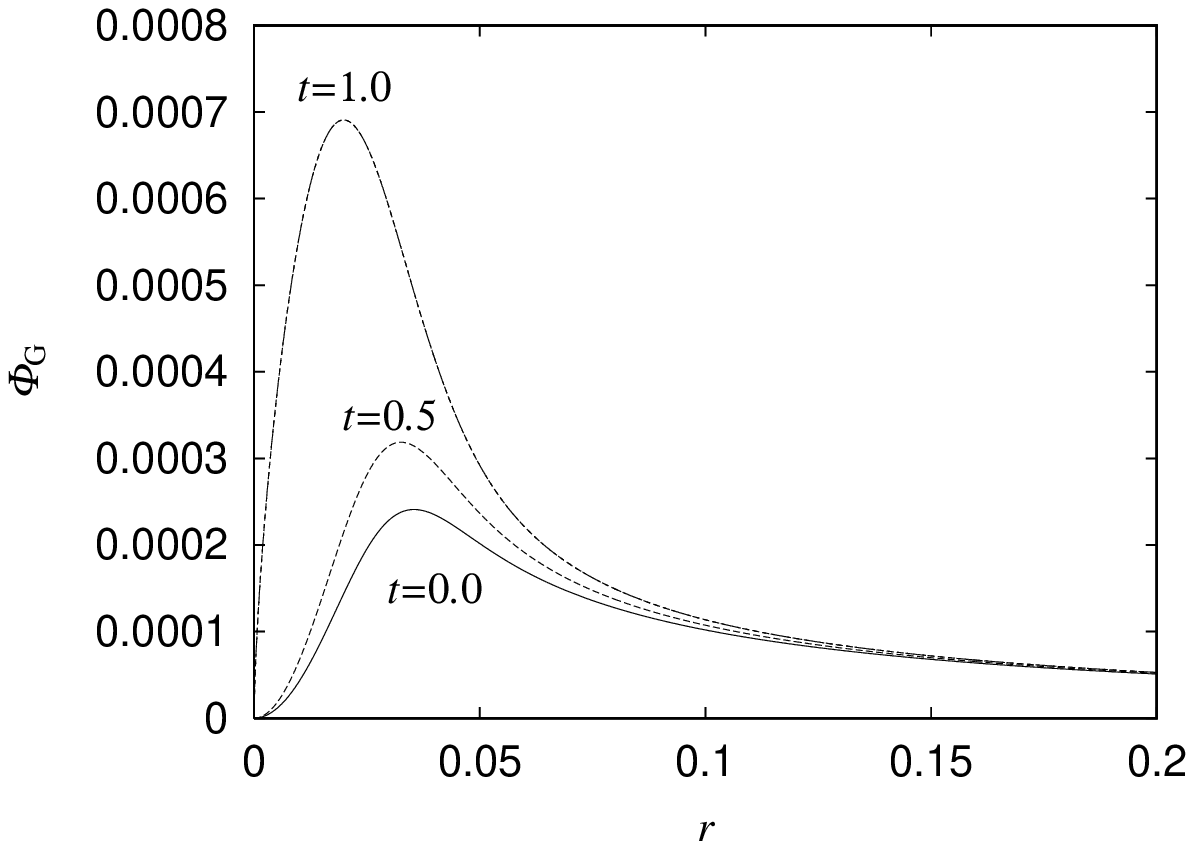}}
        \caption{The spatial configurations of the 
gravitational potential $\Phi_{\rm G}$ at $t=0$, $0.5$ and at 
the moment of the central shell focusing singularity formation $t=1$ 
are plotted. 
In all the cases, 
the gravitational potential $\Phi_{\rm G}$ is everywhere much smaller
than unity.}
\label{fg:G-Potential}
\end{figure}

\newpage

\begin{figure}
        \centerline{\epsfxsize 12cm \epsfysize 9cm \epsfbox{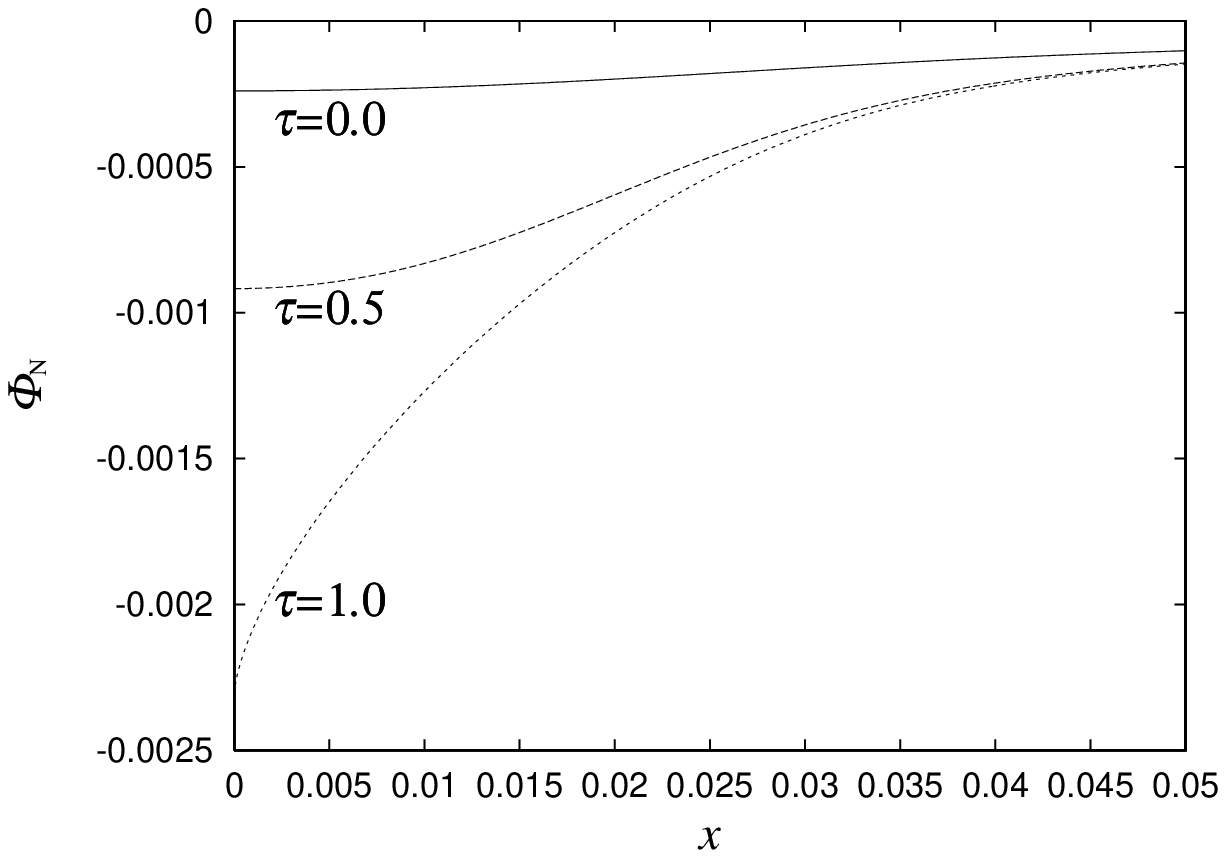}}
~(a)
\end{figure}
\begin{figure}
        \centerline{\epsfxsize 12cm \epsfysize 9cm \epsfbox{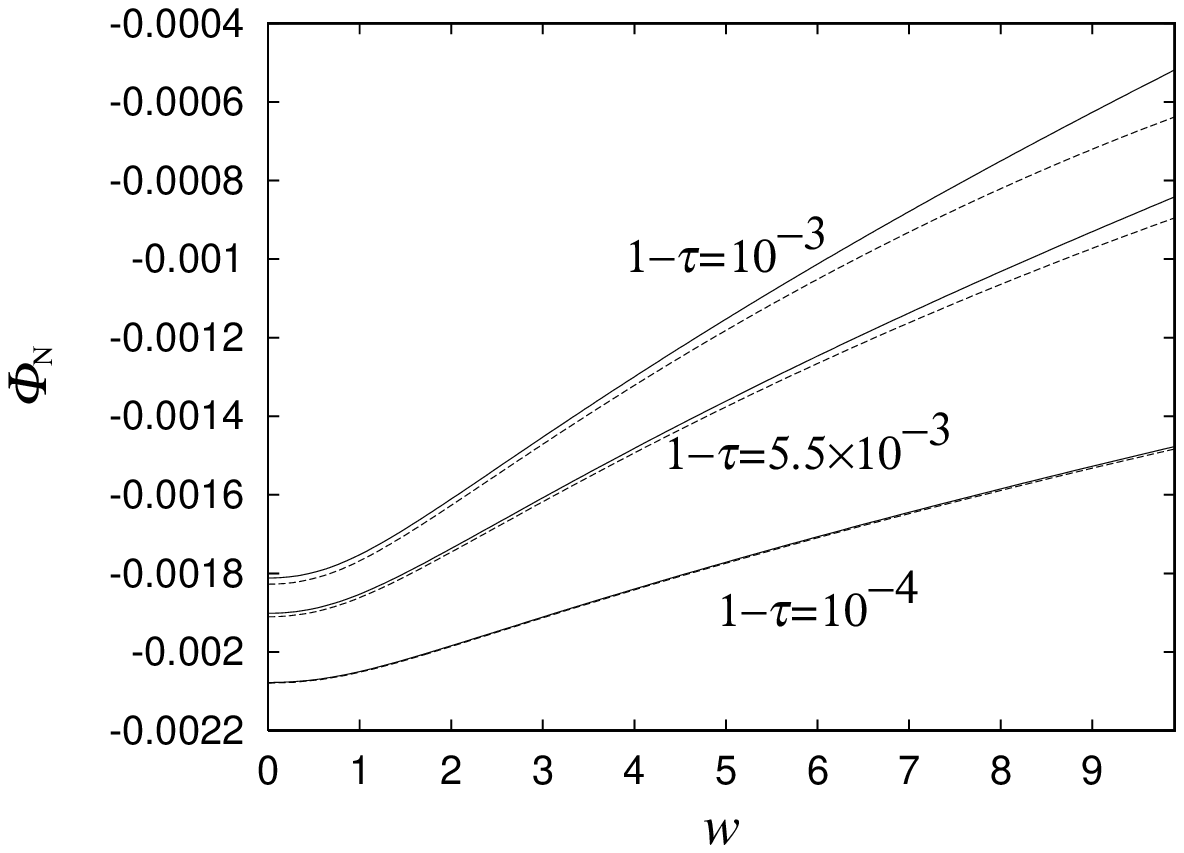}}
~(b)
        \caption{(a) The spatial configurations of the 
numerically obtained Newtonian gravitational potential $\Phi_{\rm N}$ 
are plotted as a function of 
$x$. The solid line represents $\Phi_{\rm N}$ at $\tau=0$, while 
the dashed line corresponds to the case at $\tau=0.95$. 
The dotted line represents the case at the central shell focusing 
singularity formation $t=1$. Although the regularity of $\Phi_{\rm N}$ 
at the origin breaks down at $t=1$, $|\Phi_{\rm N}|$ is everywhere 
much smaller than unity. 
 (b) The asymptotic behavior of the Newtonian gravitational potential 
$\Phi_{\rm N}$ is plotted as a function of $w$ for $\tau=0.999$ (the top), 
$\tau=0.99945$ (the middle) and $\tau=0.9999$ (the lowest).
The solid lines represent $\Phi_{\rm N}$ 
obtained by the asymptotic analysis (\ref{eq:N-Phi-ap}) while 
dashed lines is the numerically obtained ones. 
In all the cases, numerically obtained values well agree with 
those obtained by the asymptotic analysis. 
}
\label{fg:N-Phi}
\end{figure}

\newpage
\begin{figure}
        \centerline{\epsfxsize 13cm \epsfysize 10cm \epsfbox{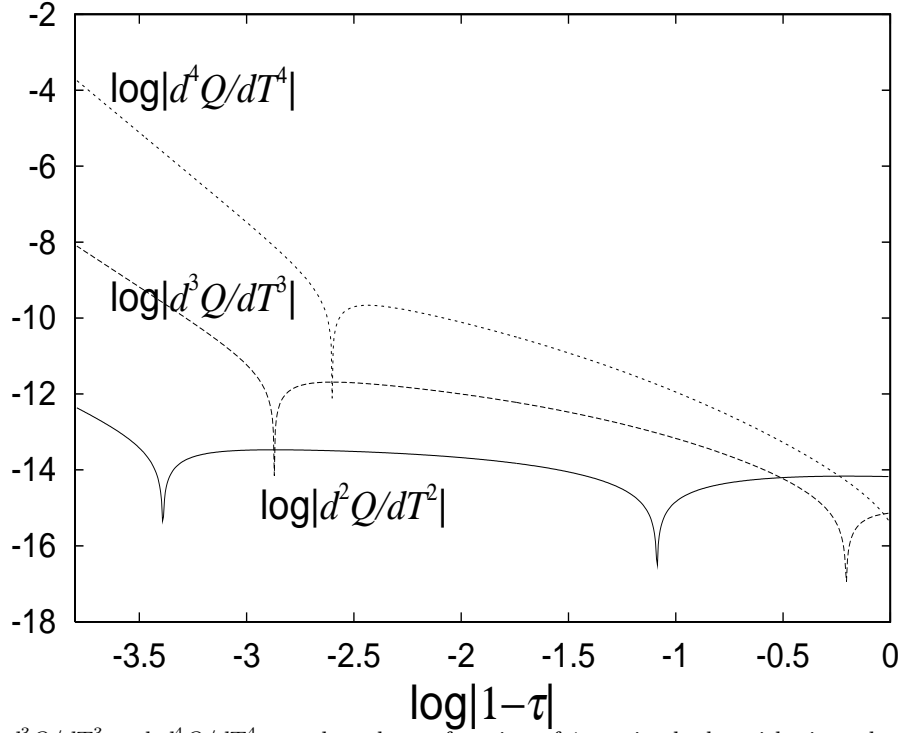}}
        \caption{$d^{2}Q/dT^{2}$, $d^{3}Q/dT^{3}$ and $d^{4}Q/dT^{4}$ 
are plotted as a function of $1-\tau$ in the logarithmic scales. 
The absolute values of these quantities show oscillatory behavior 
and grow. At this stage, the contribution near the origin 
is not dominant.}
\label{fg:Q-moment}
\end{figure}
\begin{figure}
        \centerline{\epsfxsize 13cm \epsfysize 10cm \epsfbox{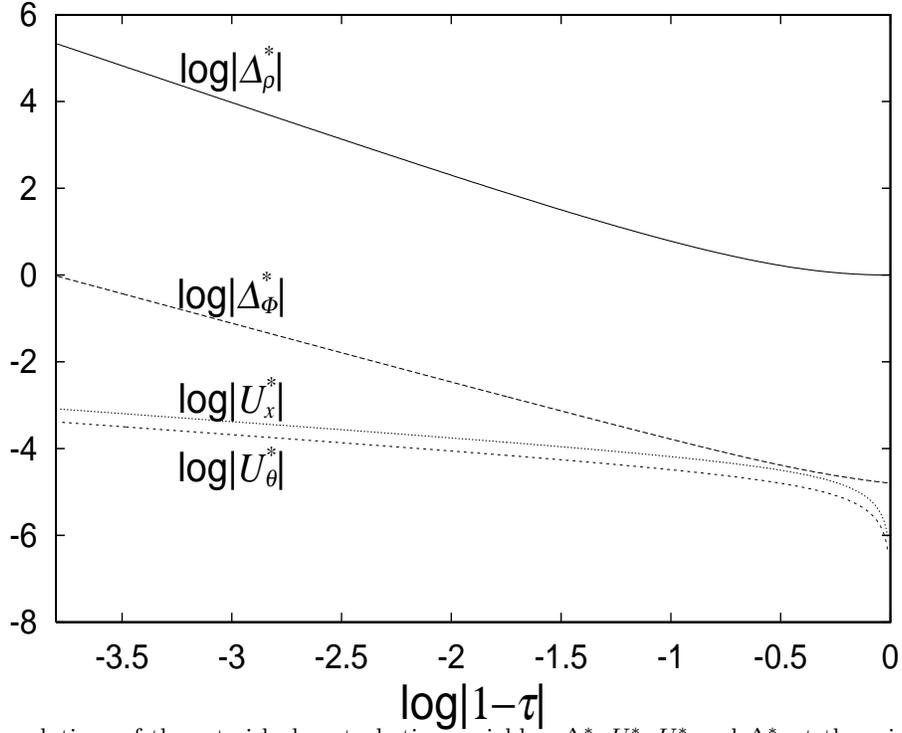}}
        \caption{The time evolutions of the asterisked perturbation 
variables, ${\Delta}^{*}_{\rho}$, ${U}^{*}_{x}$, ${U}^{*}_{\theta}$ 
and ${\Delta}^{*}_{\Phi}$ at the origin 
are plotted in the logarithmic scales. 
All these quantities monotonically grow 
asymptotically in accordance with the power low. }
\label{fg:hatted-variable}
\end{figure}

\newpage
\begin{figure}
        \centerline{\epsfxsize 13cm \epsfysize 9cm \epsfbox{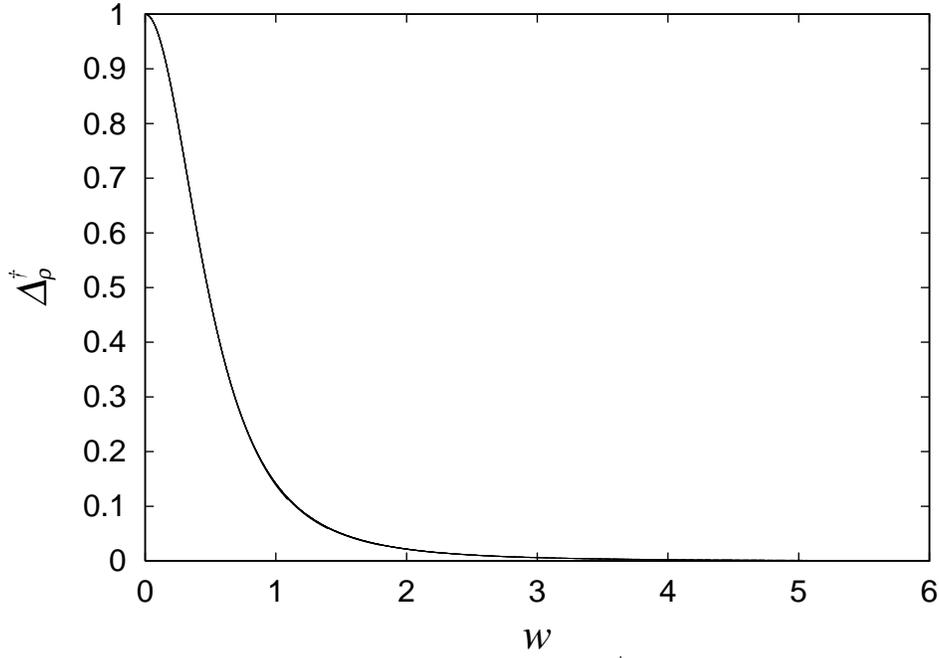}}
        \caption{Spatial configurations of the 
normalized density perturbation ${\Delta}^{\dagger}_{\rho}$ are plotted 
as a function of $w$ for $x<0.05x_{\rm core}$ 
at various time step form $1-\tau=6.6\times10^{-3}$ 
to $1-\tau=2.0\times10^{-4}$ by solid lines. 
The density perturbation becomes about 381 times during this time
 interval. 
On the other hand, ${\delta}^{*}_{\rho}(w)/\delta_{(0)}$ obtained by the 
asymptotic estimate is represented by a dashed line. 
It is very difficult to distinguish these lines from each other. } 
\label{fg:normarized-Drho}
\end{figure}
\begin{figure}
        \centerline{\epsfxsize 13cm \epsfysize 9cm \epsfbox{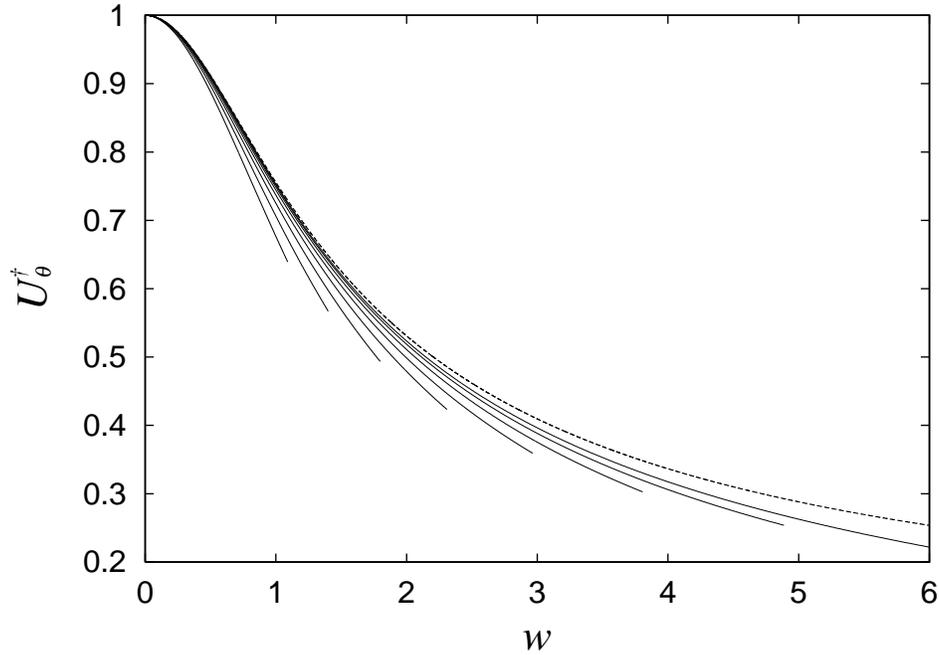}}
        \caption{Spatial configurations of the 
normalized tangential velocity perturbation ${U}^{\dagger}_{\theta}$ 
are plotted 
as a function of $w$ for $x<0.05x_{\rm core}$ at various time step form $1-\tau=6.6\times10^{-3}$ 
to $1-\tau=2.0\times10^{-4}$ by solid lines. 
On the other hand, ${u}^{*}_{\theta}(w)$ obtained by the 
asymptotic estimate is represented by a dashed line. 
We can see that ${U}^{\dagger}_{\theta}$ asymptotically approaches to 
${u}^{*}_{\theta}$. } 
\label{fg:normarized-Ut}
\end{figure}

\newpage
\begin{figure}
        \centerline{\epsfxsize 13cm \epsfysize 9cm \epsfbox{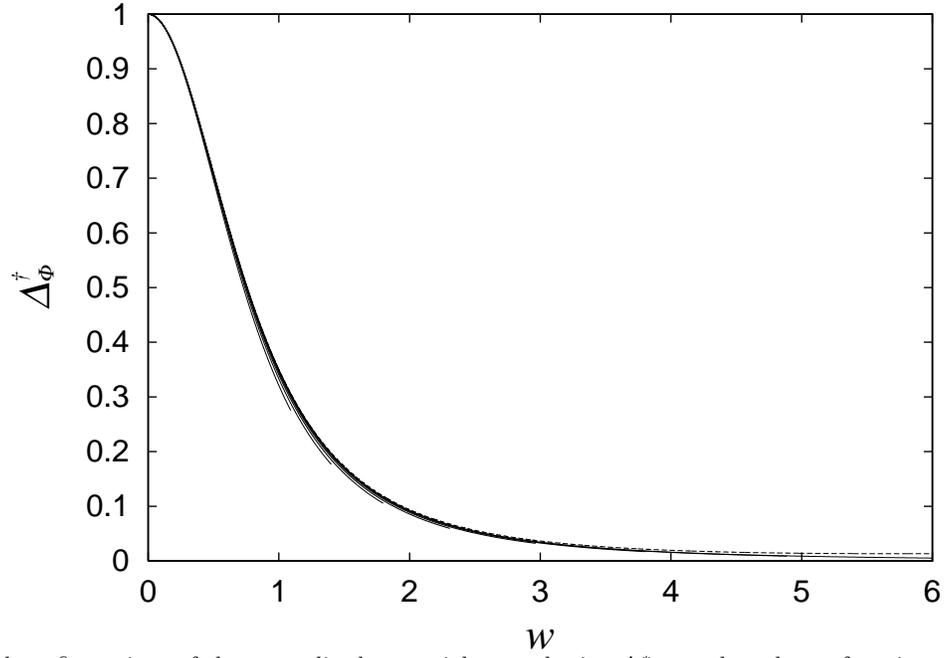}}
        \caption{Spatial configurations of the 
normalized potential perturbation ${\Delta}^{*}_{\Phi}$ are plotted 
as a function of $w$ for $x<0.05x_{\rm core}$ 
at various time step form $1-\tau=6.6\times10^{-3}$ 
to $1-\tau=2.0\times10^{-4}$ by solid lines. 
On the other hand, ${\delta}^{*}_{\Phi}(w)/\Delta_{(0)}$ obtained by the 
asymptotic estimate is represented by a dashed line. 
The normalized potential perturbation $\Delta^{\dagger}_{\Phi}$ 
asymptotically approaches to $\delta^{*}_{\rho}/\delta_{(0)}$. }
\label{fg:normarized-Dphi}
\end{figure}

\end{document}